\documentclass[12pt]{article}
\usepackage{amssymb}
\usepackage{amsmath}
\usepackage{cprotect}
\usepackage{tikz}
\usepackage{cite}
\usepackage{youngtab}
\usepackage{hyperref}
\input xy
\xyoption{all}
\usepackage{enumitem}

\def\sqr#1#2{{\vcenter{\vbox{\hrule height.#2pt
            \hbox{\vrule width.#2pt height#1pt \kern#1pt
                  \vrule width.#2pt}\hrule height.#2pt}}}}

\def\sqra#1#2#3{{\vcenter{\vbox{\hrule height.#2pt
            \hbox{\vrule width.#2pt height#1pt \kern5pt 
#3
                  \vrule width.#2pt}\hrule height.#2pt}}}}

\numberwithin{equation}{section}

\numberwithin{table}{section}

\begin{document}

\begin{center}

{\large\bf Chern-Simons theory, decomposition, and the A model}

\vspace*{0.2in}

Tony Pantev$^1$, Eric Sharpe$^2$, Xingyang Yu$^2$

        \vspace*{0.1in}
        
        \begin{tabular}{cc}
                {\begin{tabular}{l}
                $^1$ Department of Mathematics\\
                209 South 33rd Street\\
                Philadelphia, PA 19104-6395
                \end{tabular}}
                &
                {\begin{tabular}{l}
                $^2$ Department of Physics MC 0435\\
                                850 West Campus Drive\\
                                Virginia Tech\\
                                Blacksburg, VA  24061 \end{tabular}}
        \end{tabular}

        \vspace*{0.2in}

{\tt tpantev@math.upenn.edu},
{\tt ersharpe@vt.edu},
{\tt xingyangy@vt.edu}

\end{center}

In this paper, we discuss how gauging  one-form symmetries in Chern-Simons
theories is implemented in an A-twisted topological open string theory.  For example, the contribution from
a fixed $H/Z$ bundle on a three-manifold $M$, arising in a $BZ$ gauging of $H$ Chern-Simons, for $Z$ a finite subgroup of the center of $H$, is described by an open string worldsheet theory whose bulk is a sigma model with target a $Z$-gerbe (a bundle of one-form symmetries) over $T^*M$, of characteristic class determined by the $H/Z$ bundle.  We give a worldsheet picture of the decomposition of one-form-symmetry-gauged Chern-Simons in three dimensions, and we describe how a target-space constraint on bundles arising in the gauged Chern-Simons theory has a natural worldsheet realization.  Our proposal provides examples of the expected correspondence between worldsheet global higher-form symmetries, and target-space gauged higher-form symmetries.

\begin{flushleft}
June 2024
\end{flushleft}

\newpage

\tableofcontents

\newpage

\section{Introduction}

Years ago, the paper \cite{Witten:1992fb} argued that
classical open string field theory of the topological A model with target 
a Calabi-Yau threefold $X$, is Chern-Simons gauge theory on
a special Lagrangian submanifold $M \subset X$, or put another way,
that the D-brane worldvolume theory of the topological A model is a Chern-Simons theory.

In this paper, we shall extend \cite{Witten:1992fb} to give a worldsheet description of gauging one-form symmetries in Chern-Simons theories and their decomposition,
as discussed (in the context of Chern-Simons theories) in
\cite{Pantev:2022pbf}, for Chern-Simons theories with gauge groups realizable by open strings.

Briefly, decomposition is the observation that a $d$-dimensional
quantum field theory with a global $(d-1)$-form symmetry is equivalent
to (``decomposes into'') a disjoint union of other quantum field theories. The topological local operators generating the $(d-1)$-form symmetry build projection operators, which decompose the partition function and other observables of the theory.
This was first discussed in \cite{Hellerman:2006zs} (as part of efforts to resolve some apparent difficulties in string propagation on stacks \cite{Pantev:2005rh,Pantev:2005wj,Pantev:2005zs}) and has since been
discussed in many other places since, see for example
\cite{Anderson:2013sia,Sharpe:2019ddn,Cherman:2020cvw,Nguyen:2021yld,Nguyen:2021naa,Lin:2022xod,Perez-Lona:2023llv,Sharpe:2024ujm} for a sample of other works on the subject.
Some reviews are in
\cite{Sharpe:2010zz,Sharpe:2010iv,Sharpe:2019yag,Sharpe:2022ene}.

Compared to ubiquitous results in 2$d$ theories, decomposition in theories in dimension $d>2$ is less studied. For $3d$ Chern-Simons theories, the Chern-Simons decomposition of \cite{Pantev:2022pbf} is described as follows.
Given a Chern-Simons theory with gauge group $H$, and an action of\footnote{This is a one-form symmetry action with group $A$.} $BA$,
where $A$ is finite and abelian,
$d: A \rightarrow H$ has image in the center of $H$,
with kernel $K$ and cokernel $G$,
\begin{equation} \label{eq:decomp-ext}
1 \: \longrightarrow \: K \: \longrightarrow \: A \:
\stackrel{d}{\longrightarrow} \: H \: \longrightarrow \: G 
\: \longrightarrow \: 1,
\end{equation}
it was argued in \cite{Pantev:2022pbf} that
\begin{equation}  \label{eq:earlier-decomp}
\left[ \mbox{Chern-Simons}(H) / BA \right] \: = \:
\prod_{\theta \in \hat{K}} 
\mbox{Chern-Simons}(G)_{\theta},
\end{equation}
where the subscript $\theta$ indicates a three-dimensional discrete theta
angle. The $BA$ quotient above can be understood as gauging a one-form symmetry of the Chern-Simons theory, with $\hat{K}$ the quantum symmetry group dual to $K$. The theta angle couples to $\phi^* \omega_3$, for
$\omega_3  \in H^3_{\rm group}(G,K) = H^3_{\rm sing}(BG,K)$ that
we describe momentarily,
and
$\phi: M \rightarrow BG$ defines any given $G$ bundle on $M$.
(Both sides of the equation above have $\omega_3 = 0$; on the right,
this is enforced via the decomposition, which projects out contributions with
$\omega_3 \neq 0$.)

The discrete theta angle couples \cite{Pantev:2022pbf} to $\omega_3 = \beta_{\alpha}(w_G)$,
where $w_G \in H^2_{\rm sing}(BG,Z)$ is the obstruction to lifting $G = H/Z$ bundles to $H$ bundles for $Z = {\rm im}\, d \subset H$,
$\alpha$ the class of the extension
\begin{equation}\label{eq: A from K and Z}
1 \: \longrightarrow \: K \: \longrightarrow \: A \:
\stackrel{d}{\longrightarrow} Z \: \longrightarrow \: 1,
\end{equation}
and $\beta_{\alpha}$ the corresponding Bockstein homomorphism.
In particular, it was argued in \cite{Pantev:2022pbf} that $G$ bundles appearing in the $BA$
quotient on the left-hand-side of equation~(\ref{eq:earlier-decomp}) all have
$\beta_{\alpha}(w_G) = 0$.  On the right-hand-side of equation~(\ref{eq:earlier-decomp}), the constraint $\beta_{\alpha}(w_G) = 0$ is enforced by the decomposition: because of the varying discrete theta angles, any contributions from $G$ bundles with $\beta_{\alpha}(w_G) \neq 0$ cancel out of the path integral.

Realizing the Chern-Simons theory as a string field theory via the topological A model as in \cite{Witten:1992fb}, we propose the following worldsheet description for the gauged Chern-Simons construction above. Fix a $G$ bundle $V$ (as in any event the open string theory can only describe one $G$ bundle at a time), and let $w_G \in H^2(M,Z)$ be the obstruction to lifting $V$ to an $H$ bundle.
We claim that gauging one-form symmetries for Chern-Simons reviewed above \cite{Pantev:2022pbf} is, on the worldsheet, the topological A-twist of a sigma model with target an $A$-gerbe\footnote{
A gerbe is a stack, a close analogue of a space, which has the structure of a fiber bundle in which the fibers are (higher) groups of one-form symmetries.  A sigma model with target a $Z$-gerbe for any group $Z$ is realizable in the UV as a gauge theory in which the gauge group has a subgroup $Z$ which acts trivially \cite{Pantev:2005rh,Pantev:2005wj,Pantev:2005zs}, as we shall review.
} over $X$, which projects to a $Z$-gerbe whose restriction to $M$ has characteristic
class $w_G$, and in which the action of $A$ along the worldsheet boundary has been gauged.  The Chan-Paton factors describe (before gauging) an $H$ bundle over the restriction of the $A$-gerbe to $M$, which is equivalent to a twisted $H$ bundle over $M$, twisted by $w_G$.
(In a sigma model with target space an $A$-gerbe, implicitly, one gauges $A$ which  trivially acts on the
bulk degrees of freedom.)

We emphasize that although gauged Chern-Simons theories were described in \cite{Pantev:2022pbf} for arbitrary Lie groups, the proposed worldsheet realization is only for those Lie groups which can be described by open strings.  For example, we are not claiming to give a worldsheet description of Chern-Simons theories with exceptional gauge groups.

\subsection*{Organization of the paper}
We begin in section~\ref{sect:su2-eff} by describing the open string worldsheet theory describing
the (effective) $BZ$ gauging of the one-form symmetry in Chern-Simons with gauge group $H$, using $SU(2) / B {\mathbb Z}_2$ (for the central ${\mathbb Z}_2$) as a prototype. The starting setup is a Chern-Simons theory with gauge group $H$ as an open string field theory. The full target space is denoted as $X$ while the Chern-Simons theory is supported on a Lagrangian submanifold $M\subset X$. In the original worldsheet description of an $H$ Chern-Simons theory in \cite{Witten:1992fb}, there was the following (typical) correspondence
\begin{equation}
\begin{split}
	H~\text{global symmetry on}~&\text{the worldsheet boundary}\\
	 &\updownarrow \\
	 H~\text{gauge symmetry in the}~&\text{target space Chern-Simons}.
\end{split}
\end{equation}
As one would expect from this correspondence, the worldsheet manipulation associated to gauging the one-form $BZ$ symmetry in the Chern-Simons theory involves two key ingredients:
\begin{itemize}
    \item Gauge the $Z\subset H$ global symmetry subgroup along the worldsheet boundary so as to realize a $G=H/Z$ gauge symmetry group in the target space Chern-Simons theory.

    \item However, not all $G$ bundles can be obtained from a $Z$ quotient of an honest $H$ bundle on $M$ -- not all $G$ bundles on $M$ lift to $H$ bundles on $M$.  To describe the other $G$ bundles, the worldsheet bulk must describe a sigma model into a $Z$-gerbe over the target space manifold $X$, whose restriction to $M\subset X$ has characteristic class $w_G \in H^2(M,Z)$ (the obstruction to lifting a $G$ bundle to an $H$ bundle).\footnote{Such gerbes, from the worldsheet perspective, are described locally by gauging a $Z$ zero-form symmetry, which is trivially acting on the worldsheet bulk.  However, since the $Z$ symmetry acts non-trivially on the worldsheet boundary, this gauging does not result in a decomposition.}  All $G$ bundles on $M$ lift to an $H$ bundle on a $Z$-gerbe over $M$.  The Chan-Paton factors describe such a lift, an honest $H$ bundle on the gerbe on $M$ (which can also be interpreted as a $Z$-twisted bundle on $M$), which after the boundary gauging, projects to the desired $G$ bundle on $M$.  All $G$ bundles on $M$ can be described in this fashion.
\end{itemize}

Next, in section~\ref{sect:su2-triv}, we turn to the opposite case, of the worldsheet theory for a Chern-Simons theory in which one gauges a completely trivially-acting one-form symmetry in the target space, using
$SU(2)/B {\mathbb Z}_2$ (for a trivially-acting ${\mathbb Z}_2$) as a prototype.  We argue that this is realized on the worldsheet by gauging a trivially-acting zero-form symmetry $K=\mathbb{Z}_2$ in the bulk and on the boundary,
which, as it acts trivially on both, results in a decomposition of the worldsheet theory, reproducing the target-space decomposition. As that decomposition reflects a global one-form symmetry in the  two-dimensional theory, there is the following correspondence between symmetries on worldsheet and target:
\begin{equation}
	\begin{split}
	\text{one-form global symmetry on}~&\text{the worldsheet (decomposition)}\\
	 &\updownarrow \\
	\text{one-form gauge symmetry in} ~&\text{the target space Chern-Simons}.
\end{split}
\end{equation}

In section~\ref{sect:genl}, we make our proposal for the general\footnote{
That said, more general groupoid gaugings than those described in \cite{Pantev:2022pbf} may exist, see for example \cite{Perez-Lona:2023llv} in a different context, but here our focus is on giving a worldsheet realization of the construction of \cite{Pantev:2022pbf}.
} case of \cite{Pantev:2022pbf}, combining the two special cases above. Namely, instead of just gauging the center $Z$ or gauging the trivially acting $K$, we gauge the group $A$ whose connection to $Z$ and $K$ is shown in~(\ref{eq: A from K and Z}). Here, the worldsheet bulk is a sigma model into an $A$-gerbe over the target space manifold $X$, lifting the $Z$-gerbe needed to describe a specific $G = H/Z$ bundle.  As we discuss there, such lifts only exist when 
$\beta_{\alpha}(w_G) = 0$.  This reproduces on the worldsheet the property of the $BA$-gauged Chern-Simons target-space
theory utilized in \cite{Pantev:2022pbf}, and serves as an important consistency check on our construction.  We also discuss the decomposition from the worldsheet perspective.

In section~\ref{sect:su2-z4} we briefly discuss a prototypical example of the general case,
namely gauging a ${\mathbb Z}_4$ symmetry of the $SU(2)$ Chern-Simons theory, which projects to the central $Z={\mathbb Z}_2$ of $SU(2)$, with a $K={\mathbb Z}_2$ kernel.  This was one of the prototypical examples of
\cite{Pantev:2022pbf}.

In section~\ref{sect:undoing} we discuss the worldsheet realization of the target-space manipulation of gauging the global two-form symmetry, which `undoes' the decomposition, and selects out a single universe.\footnote{Although not rigorous, one can intuitively understand the quantum symmetry of this gauging as a `(-1)-form symmetry,' whose background field is the parameter labeling universes.}
Finally, in section~\ref{sect:bmodel} we discuss analogous considerations in topologically B-twisted sigma models,
and implications for gauged holomorphic Chern-Simons theories in six real dimensions.

In appendix~\ref{app:revcs} we briefly review how Chern-Simons theory is the open string field theory of the A model, following \cite{Witten:1992fb}.
In appendix~\ref{app:global} we briefly review which Lie groups can arise in
worldsheet boundary constructions, and in appendix~\ref{app:gauge-qm} we make some technical observations on gauging finite symmetries in quantum mechanics.

In passing, we should note that away from the large-radius limit,
the Chern-Simons theory will be deformed by worldsheet instanton
corrections.  In this paper we implicitly restrict to the large-radius
limit -- we will not analyze those open string worldsheet instanton corrections here.

\section{Prototype:  $SO(3)$ from $SU(2)/{\rm center}$}
\label{sect:su2-eff}

Before getting to the worldsheet realization of Chern-Simons$(H)/BA$ for general (realizable)
$H$ and $A$, let us first examine, formally, the worldsheet realization of the special case of $H = SU(2)$ Chern-Simons theories
with
$A = {\mathbb Z}_2$ which maps directly to the center $Z=\mathbb{Z}_2 \subset SU(2)$.  This gauge group is realizable\footnote{
More precisely, its Lie algebra is realizable, and we leave questions about existence of theories with specific global gauge group forms for later work.
} in open strings as $SU(2) = USp(2)$ (see appendix~\ref{app:global}).  The result of the target-space gauging is a single copy of Chern-Simons theory with the gauge group $G=H/Z=SO(3)$.  The worldsheet realization of gauging $H$ Chern-Simons by $BZ$ for finite effectively-acting $Z$ is a simple generalization, as we also discuss.

Much as in a heterotic string sigma model, a fixed worldsheet theory should describe a fixed $SO(3)$ bundle on the target space $M$, which we denote $V$.  (Then, the string field theory will (nonperturbatively) sum over bundles, to reproduce Chern-Simons theory.)  
Let $w_2$ denote the second Stiefel-Whitney class of $V$, which is
the obstruction to lifting $V$ to an $SU(2)$ bundle denoted by $V_\omega$.  We should distinguish the following two cases:
\begin{itemize}
    \item If $w_2 = 0$, then the $SO(3)$ bundle can be lifted to an $SU(2)$ bundle, meaning that there exists an $SU(2)$ bundle $V_w \rightarrow M$ such that
$V = V_w/{\mathbb Z}_2$ is the desired $SO(3)$ bundle.  Conversely, any principal $SU(2)$ bundle on $M$ automatically has at least one\footnote{Choices of ${\mathbb Z}_2$ equivariant structure on an $SU(2)$ bundle, when the ${\mathbb Z}_2$ acts trivially on the base, are just choices of ${\mathbb Z}_2$ action on $SU(2)$.  By identifying the ${\mathbb Z}_2$ with the center of $SU(2)$, we are implicitly giving one equivariant structure.} 
action of ${\mathbb Z}_2$, known technically as a ${\mathbb Z}_2$-equivariant structure.
\item If on the other hand $w_2 \neq 0$, there is no honest $SU(2)$ bundle $V_w \rightarrow M$ such that $V_w/{\mathbb Z}_2$ is the desired $SO(3)$ bundle.  However, one can construct a twisted $SU(2)$ bundle $V_w$ over $M$, or equivalently an honest $SU(2)$ bundle over a ${\mathbb Z}_2$ gerbe over $M$, of characteristic
class $w_2$, such that $V = V_w/{\mathbb Z}_2$, realizing the $SO(3)$ bundle $V$.

Let us describe this case more explicitly.  Let $g_{\alpha \beta}$ be transition functions for an $SO(3)$ bundle, and let $\tilde{g}_{\alpha \beta}$ be a set of lifts to $SU(2)$ on intersections of open patches.  (We assume that the intersections are small enough that lifts always exist locally, though there can be obstructions to a global lift.)
Since $SU(2)$ is a double cover of $SO(3)$, there are two choices of $\tilde{g}_{\alpha \beta}$ for any one $g_{\alpha \beta}$.  Now, the transition functions of the original $SO(3)$ bundle close on triple overlaps,
by definition of bundle:
\begin{equation}
    g_{\alpha \beta} g_{\beta \gamma} g_{\gamma \alpha} \: = \: 1.
\end{equation}
However, there is no guarantee that the lifts will close.  In general,
we can merely write
\begin{equation} \label{eq:twist-defn}
    \tilde{g}_{\alpha \beta} \tilde{g}_{\beta \gamma} \tilde{g}_{\gamma \alpha} \: = \: h_{\alpha \beta \gamma},
\end{equation}
for some $h_{\alpha \beta \gamma} \in {\mathbb Z}_2$.  It is easy to check that
the $h_{\alpha \beta \gamma}$ is closed on quadruple overlaps, and so defines an element of $H^2(M,{\mathbb Z}_2)$, which in this case is precisely the second Stiefel-Whitney class (and more generally, will be the cohomology class giving the obstruction to lifting a $G = H/Z$ bundle to an $H$ bundle).  We refer to a bundle whose transition functions only close on triple overlaps as 
in~(\ref{eq:twist-defn}) as a twisted bundle.

Although a twisted bundle would not arise from Chan-Paton factors over a target space submanifold $M\subset X$,
such a Chan-Paton bundle will
arise if\footnote{
A more familiar example to the reader may be the relationship between nontrivial $B$ fields and twists of Chan-Paton factors, see e.g.~\cite[section 5.3]{Witten:1998cd}.  Specifically, as noted in for example \cite[equ'n (1.8)]{Freed:1999vc}, under a gauge transformation $B \mapsto B + d \Lambda$, the Chan-Paton gauge
field $A$ necessarily undergoes $A \mapsto A - \Lambda$.  This means that if the $B$ field is topologically nontrivial, then the Chan-Paton bundle is twisted.  Now, topologically nontrivial $B$ fields on $M$ are characterized by (torsional elements of) $H^3(M,{\mathbb Z})$ and $H^2(M,U(1))$, whereas here, we want $H^2(M,Z)$ for finite $Z$, which need not be related.  As a result, the twists we need are not necessarily describable by topologically nontrivial $B$ fields, but can always be described by gerbes.  See also \cite{Anderson:2013sia} for an analogous discussion of twisted bundles in heterotic strings on gerbes.
} the target space is a ${\mathbb Z}_2$ gerbe over $M$ (essentially, a fiber bundle over $M$ with fibers $B {\mathbb Z}_2$).  Phrased another way, there are more bundles on a gerbe over $M$, than there are bundles just on $M$ -- we can pullback a bundle on $M$ to get a bundle on $M$, and in addition, there are also bundles on the gerbe which do not arise by pullback from $M$.  The bundles on the gerbe which are not pullbacks, are honest bundles on the gerbe, and can also be equivalently interpreted as twisted bundles on the space $M$, twisted in the sense above.  We will utilize that fact in our construction, and so will discuss constructions of sigma models with gerbe target spaces momentarily.
\end{itemize}

Now, let us consider how to realize this structure on the worldsheet.  Since the worldsheet bulk and boundary
theories are separate theories (albeit linked) with different  path integrals, one can distinguish gauging along the boundary from gauging in the bulk. Physically, this can be done via summing over (i.e., condensing) topological operators just along the boundary or in the full worldsheet bulk.
Briefly, we have the following two basic gauging manipulations as building blocks.
\begin{itemize}
    \item Gauging the ${\mathbb Z}_2$ along the boundary will project the zero-form worldsheet global symmetry $SU(2)$ to $SO(3)$, hence changing the target-space gauge group from $SU(2)$ to $SO(3)$. As discussed in appendix~\ref{app:gauge-qm}, such a gauging inserts boundary operators corresponding to
    elements of ${\mathbb Z}_2$.  The boundary partition function, a simple statistical mechanical theory, gets an insertion of $1+z$, which projects onto ${\mathbb Z}_2$-invariant states,
    and so implements the projection onto $SO(3)$ degrees of freedom.  That is clearly part of what is needed -- however, as observed above, unless the $SO(3)$ bundle has vanishing $w_2$, it will not suffice to describe a general
    $SO(3)$ bundle.
    \item Gauging the ${\mathbb Z}_2$ in the bulk, which is trivially acting no bulk degrees of freedom, implements a ${\mathbb Z}_2$ gerbe, whose characteristic class will be the $w_2$ of the desired $SO(3)$ bundle.  The boundary
    theory then describes ${\mathbb Z}_2$-twisted $SU(2)$ bundles, which is part of what we need for our construction. 
    
    Constructions of sigma models with gerbes as target spaces were discussed in \cite{Pantev:2005rh,Pantev:2005wj,Pantev:2005zs}.  For example, a nonlinear sigma model with target a ${\mathbb Z}_k$ gerbe over $M$ can be constructed by starting with
a sigma model with target the total space of a $U(1)$ bundle
$L \rightarrow M$, and then gauging by $k$ rotations of $U(1)$, so that a trivially-acting ${\mathbb Z}_k \subset U(1)$ remains.  The resulting theory has a global $B {\mathbb Z}_k$ symmetry (reflecting translations along the fibers of the gerbe), and describes a gerbe of characteristic class $c_1(L) \mod k \in H^2(M,{\mathbb Z}_k)$.
\end{itemize}

Combining the above two choices, we now have several options of gauging on the worldsheet theory:
\begin{itemize}
    \item One option is to only gauge the ${\mathbb Z}_2$ along the boundary theory describing
    $SU(2)$, but not gauge the trivial action of ${\mathbb Z}_2$ in the bulk. If one realizes the $\mathbb{Z}_2$ symmetry via topological line operators on the worldsheet, this means summing over topological lines only along the boundary. However, as we explained, this could never realize a nontrivial ${\mathbb Z}_2$ gerbe over $M$, and it could only construct $SO(3)$ bundles with a trivial second Stiefel-Whitney class $w_2 = 0$, since the starting point would be an honest $SU(2)$ bundle on $M$.  As already described, this is not the most general case, and does not describe all $SO(3)$ bundles.  
    \item Another option is to gauge the ${\mathbb Z}_2$ in the bulk, where it is trivially-acting, but not gauge it along the boundary. From the topological defect perspective, this amounts to condensing the topological line, generating $\mathbb{Z}_2$, on the whole worldsheet except for the boundary manifold. This would give rise to a ${\mathbb Z}_2$ gerbe structure over $M$, with fixed characteristic class $w_2 \in H^2(M,{\mathbb Z}_2)$, but, would not project $SU(2)$ to $SO(3)$ along the boundary, as we shall elaborate shortly.  (More generally, in e.g.~D-branes in orbifolds, one gauges the orbifold group in the bulk of the worldsheet, but not the action on the boundary Chan-Paton factors.)
    \item In order to get both a nontrivial ${\mathbb Z}_2$ gerbe structure over $M$ and to project $SU(2)$ to $SO(3)$, as needed to describe $SO(3)$ bundles on $M$ of $w_2 \neq 0$, we must gauge the ${\mathbb Z}_2$ in both bulk and boundary.  This is our proposal, as we shall elaborate below.
\end{itemize}

To summarize, our proposal is as follows.
Fix an $SO(3)$ bundle $V$ on $M$, and let $w_2 \in H^2(M, {\mathbb Z}_2)$
be the obstruction to lifting the $SO(3)$ bundle $V$ to an $SU(2)$ bundle (the second Stiefel-Whitney class).  Let $M_w$ be a ${\mathbb Z}_2$ gerbe on $M$ with characteristic class $w_2$, and let
$X_w$ be its pullback to $X = T^* M$.  Let $V_w$ be an $SU(2)$ bundle on $M_w$ (equivalently, a twisted $SU(2)$ bundle on $M$) lifting $V$.
We propose that the worldsheet description of the $SO(3)$ Chern-Simons theory
is the string field theory of the A model with target $X_w$, describing a D-brane on $M_w$,
and with $SU(2)$ bundle $V_w \rightarrow M_w$.  We gauge ${\mathbb Z}_2$ in both
bulk and boundary.  The bulk ${\mathbb Z}_2$ gauging is implicit in the definition of a sigma model with target a gerbe -- the action on bulk degrees of freedom is (locally) trivial,
and encoded in the structure of the gerbe (as explained in \cite{Pantev:2005rh,Pantev:2005wj,Pantev:2005zs});
the action on the boundary is the usual action of the center of 
$SU(2)$.

This proposal is also consistent with the construction of Wilson lines along the open string boundary\footnote{
On a suitable three-manifold, such as $T^3$ or a Seifert-fibered space,
the Wilson lines detect all characteristic classes.
(On other three-manifolds, such as $S^3$, some characteristic classes
are invisible to Wilson lines.)  Thus, by restricting to suitable
three-manifolds, the Wilson lines on the open string boundaries will
detect all characteristic classes.
}.
Recall that an $SO(3)$ bundle which cannot be lifted to $SU(2)$ has (some) Wilson lines which can only locally be lifted to $SU(2)$, meaning that any lift only closes up to an insertion of an element of the center.  That insertion along the boundary is implemented as the endpoint of a branch cut emanating from a bulk twist field for a trivially-acting ${\mathbb Z}_2$, as schematically illustrated below:
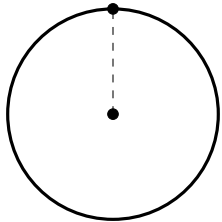
\begin{center}
\begin{tikzpicture}
    \draw[very thick] (0,0) circle (1.4);
    \draw[dashed] (0,0) -- (0,1.4);
    \filldraw[black] (0,0) circle (2pt);
    \filldraw[black] (0,1.4) circle (2pt);
\end{tikzpicture}
\end{center}
Thus, to get the Wilson lines of an $SO(3)$ theory with $w_2 \neq 0$, we need to gauge the
 ${\mathbb Z}_2$ also in bulk, though it is trivially acting on the bulk.

The effect of gauging a trivially-acting ${\mathbb Z}_2$ in the worldsheet bulk, and how that
leads to a ${\mathbb Z}_2$ gerbe structure on the target space, has been extensively discussed
in \cite{Pantev:2005rh,Pantev:2005wj,Pantev:2005zs}.  We review the effect of gauging a finite group
in a quantum mechanical system, such as the open string boundary, in appendix~\ref{app:gauge-qm}.
Briefly, the worldsheet theory for the $SO(3)$ Chern-Simons theory looks like that for the $SU(2)$ theory, but with insertions of pointlike operators along the boundary representing the action of elements of the center of $SU(2)$.

If one were to formally compute the partition function of the theory, then
we would sum over those insertions.  Formally, the boundary theory would
contain
\begin{equation}
    1 + z
\end{equation}
($z$ generating the center), which is proportional to a projector mapping
to states invariant under the ${\mathbb Z}_2$, replacing $SU(2)$ with
$SO(3)$, as expected.

Let us take a moment to summarize the relationship between various symmetries appearing here.
\begin{itemize}
    \item In the worldsheet bulk, we have a global $B {\mathbb Z}_2$ (one-form) symmetry, from the fact that we are gauging a trivially-acting ${\mathbb Z}_2$ zero-form symmetry.
    \item In the target space, we have a gauge(d) $B {\mathbb Z}_2$ one-form symmetry associated to the center of the $SU(2)$ gauge group of the Chern-Simons theory.
\end{itemize}
This naturally matches the usual correspondence between global symmetries on the worldsheet
and gauge symmetries in the target space.  Similarly, 
\begin{itemize}
    \item On the worldsheet boundary, we have a global $SO(3)$ symmetry (on Chan-Paton factors), remaining after gauging
    the ${\mathbb Z}_2$ subgroup of the global $SU(2)$ symmetry.
    \item In the target space, there is a  $SO(3)$ gauge symmetry for the Chern-Simons theory.
\end{itemize}
Again, this naturally matches the usual correspondence between global symmetries on the
worldsheet and gauge symmetries in the target space.

Next, let us understand why there is no decomposition.  Ordinarily, in two-dimensional theories, if one gauges a trivially-acting zero-form symmetry, the theory has a global one-form symmetry, and so decomposes \cite{Hellerman:2006zs}.
Here, although the bulk degrees of freedom have a gauged trivially-acting
${\mathbb Z}_2$, that same ${\mathbb Z}_2$ acts nontrivially on the boundary,
and so, the action on the entire theory is nontrivial.  As a result, there is no global one-form
symmetry, and no decomposition.  

We should note that a similar phenomenon happens in two-dimensional theories with a boundary in which one does not gauge the boundary, as in \cite{Hellerman:2006zs}.  Consider, for example, a two-dimensional orbifold in which a subgroup $K$ of the orbifold group acts trivially on the bulk degrees of freedom.  It can still act nontrivially on the boundary degrees of freedom, and as the bulk action is trivial, the boundary action (really, a choice of equivariant structure on the Chan-Paton bundle) is determined by a representation of $K$ indexing the closed-string universe.  Fixing that representation fixes a bulk universe relative to a decomposition of a two-dimensional closed string theory.  So, the closed-string theory decomposes, and for each closed-string universe there is an open string theory, with boundary action determined by the representation of $K$.  In this fashion, decomposition reproduces standard mathematics results, saying for example that the K theory of a gerbe matches the K theory of a disjoint union of spaces, as discussed in \cite{Hellerman:2006zs}.

A similar phenomenon arises in heterotic
strings on gerbes.  There, one can construct examples in which a gauge symmetry
acts trivially on right-movers, but nontrivially on left-movers -- essentially,
a gauged trivial group action on the space, but with a nontrivial action on the
bundle, and there is no decomposition.  This heterotic analogue is discussed
further in \cite{Anderson:2013sia}.

Next, let us turn to states and operators in the boundary quantum mechanics.  As discussed in appendix~\ref{app:gauge-qm}, since the spatial cross-sections of a timelike $S^1$ are pointlike, there can be no twisted sectors, no additions to the state space, but the gauging does add operators to the $SU(2)$ theory, corresponding to the elements of the group ${\mathbb Z}_2$.  These operators generate branch cuts in the $SU(2)$ theory -- they implement the path integral's sum over bundles.

In terms of Wilson lines around the boundary of the open worldsheet disks, we are taking advantage of the fact that an $SO(3)$ Wilson line can be described as the sum of two possible $SU(2)$ Wilson line lifts:  one with the identity inserted, and one with the generator of the ${\mathbb Z}_2$ center inserted.  (The vev of one or the other $SU(2)$ Wilson line may vanish, depending upon the $SO(3)$ Wilson line.)  This corresponds to the fact that the effect of the gauging is to insert a projection operator into the partition function for the $SU(2)$ theory.

So far we have discussed the worldsheet realization of gauging $B {\mathbb Z}_2$ in $SU(2)$ Chern-Simons theory, but this easily generalizes.  Let $H$ be any Lie group (that can be described by open string Chan-Paton factors), and $Z$ any subgroup of
the center of $H$.  Fix a $G = H/Z$ bundle, call it $V$, and let $w_G \in H^2(M,Z)$ the obstruction to lifting $V$ to an honest $H$ bundle on $M$. Let $M_w$ be a $Z$-gerbe on $M$ of characteristic class $w_G$, and $X_w$ its pullback
to $T^* M$.  Although the $G$ bundle $V$ may not lift to an $H$ bundle on $M$, it does lift to an $H$ bundle
on $M_w$, which is a special Lagrangian in $X_w$. The resulting $H$ bundle is then denoted by $V_w \rightarrow M_w$.
(This is an honest $H$ bundle on the gerbe $M_w$, and simultaneously a twisted $H$ bundle on
the underlying space $M$.)

Then, we propose the following worldsheet realization of the target-space $BZ$ gauging
of $H$ Chern-Simons.  For fixed $G = H/Z$ bundle $V$, 
we take the worldsheet bulk theory to be a sigma model with target the $Z$-gerbe $X_w$, such that
the boundary maps to the special Lagrangian $Z$-gerbe $M_w$, of characteristic class $w_G \in H^2(M,Z)$. 
The worldsheet boundary Chan-Paton factors describe the $H$ bundle $V_w \rightarrow M_w$ over the $Z$-gerbe,
and the action of $Z$ is gauged along the boundary.

Now, a sigma model with target a $Z$-gerbe is a local trivially-acting $Z$ gauge theory, or more explicitly, is described in the UV by a gauge theory where the gauge group includes trivially-acting $Z$ as a subgroup   \cite{Pantev:2005rh,Pantev:2005wj,Pantev:2005zs},  As a result, we can describe this proposal by saying that, at least locally, $Z$ is gauged on both the worldsheet bulk and boundary, and acts trivially on the bulk degrees of freedom.

  Although $Z$ acts trivially on the bulk degrees of freedom, it acts nontrivially on the boundary degrees of freedom. So, there is no decomposition, much as in the heterotic
analogs discussed in \cite{Anderson:2013sia}.

Much as in the $SU(2)$ case, one has the following global-gauge symmetry correspondence for the worldsheet and the target space theories.
\begin{itemize}
    \item The worldsheet bulk has a global $BZ$ (one-form symmetry), which corresponds to the target-space gauge(d) $BZ$ one-form symmetry.
    \item The worldsheet boundary has a global $G$ zero-form symmetry, which corresponds to the target-space $G$ gauge symmetry for the Chern-Simons gauge theory.
\end{itemize}

\section{Prototype:  $SU(2)/{\mathbb Z}_2$ for trivially-acting ${\mathbb Z}_2$}
\label{sect:su2-triv}

In this section we consider the target space theory Chern-Simons$(SU(2))/B{\mathbb Z}_2$, where the
${\mathbb Z}_2$ acts trivially\footnote{
Let us take a moment to clarify what it means for a one-form symmetry
to act `trivially,' following \cite{Pantev:2022pbf}.  Recall that one-form symmetries in 3$d$ are generated by topological line operators, charging line defects \cite{Gaiotto:2014kfa}. We say a one-form symmetry is trivially acting when no line operator is charged, no matter whether there are local operators carrying one-form symmetry charges. We encourage the
reader to consult \cite{Pantev:2022pbf} for further details.
Presumably, a trivially acting $p$-form symmetry could be defined as that no $p$-dimensional defect is charged, regardless of whether there are other dimensional charged objects. We leave this generalization and its application to decomposition for future work.}.  In other words, in the notation of \cite{Pantev:2022pbf},
$A = K = {\mathbb Z}_2$, and $d: A \rightarrow SU(2)$ maps all of $A$ to the identity.
In this case, the prediction of \cite{Pantev:2022pbf} is that this theory should be equivalent to a disjoint union of two copies of the $SU(2)$ theory:
\begin{equation}
    \left[ \mbox{Chern-Simons}(SU(2)) / B {\mathbb Z}_2 \right] \: = \:
    \coprod_{\theta \in \hat{\mathbb Z}_2} \mbox{Chern-Simons}(SU(2)).
\end{equation}
More generally, the copies will couple to different discrete theta angles, but in this simple example, those discrete theta angles vanish.

On the worldsheet, this can be implemented by gauging a trivially-acting
${\mathbb Z}_2$, in both bulk and boundary.

Gauging the trivially-acting ${\mathbb Z}_2$ in the worldsheet bulk means technically that the target is a ${\mathbb Z}_2$ gerbe over
$X$, (with boundaries restricted to a ${\mathbb Z}_2$ gerbe on $M$,) rather than $M$ or $X$ itself.  For this example, the trivial gerbe $[X/{\mathbb Z}_2]$ will suffice.
We could also consider nontrivial ${\mathbb Z}_2$ gerbes over $M$
(characterized by elements of $H^2(M,{\mathbb Z}_2)$), and for the reasons
discussed in \cite{Hellerman:2006zs}, the effect of such a gerbe is simply to 
shift the values of the $B$ fields on different universes in the decomposition, as the image under the characters of the map
\begin{equation}
    H^2(M,{\mathbb Z}_2) \: \stackrel{\theta}{\longrightarrow} \:
    H^2(M,U(1)).
\end{equation}
Let $X_z$ denote the chosen ${\mathbb Z}_2$ gerbe over $X$, and $M_z$ the restriction of that
gerbe to $M$.

If $V$ denotes the fixed $SU(2)$ bundle on $M$, then technically the boundary Chan-Paton factors live on
$V_z = \pi^* V$, where $\pi: M_Z \rightarrow M$ is the projection.  In terms of group actions,
this means that the ${\mathbb Z}_2$ acts trivially on $V$.

Since the target of the bulk theory is the gerbe $X_z$, implicitly we have gauged a trivially-acting ${\mathbb Z}_2$ in the worldsheet
bulk \cite{Pantev:2005rh,Pantev:2005wj,Pantev:2005zs}.  In other words, the theory is described as having a local trivially-acting ${\mathbb Z}_2$, and has a UV presentation as a gauge theory with a trivially-acting
${\mathbb Z}_2$ subgroup.

As observed in appendix~\ref{app:gauge-qm} and elsewhere, gauging the trivially-acting ${\mathbb Z}_2$ along the boundary does not generate any new states,
as there are no twisted sectors in quantum mechanics (since spatial cross-sections are zero-dimensional).  That said, the boundary gauging does generate
an additional operator ${\cal O}$ which implements that trivially-acting ${\mathbb Z}_2$, reflecting the fact that the boundary path integral is summing over
principal ${\mathbb Z}_2$ bundles on the boundary $S^1$, of which there are two distinct isomorphism classes.
Furthermore, this new operator commutes with all other operators (as the
${\mathbb Z}_2$ acts trivially), and ${\cal O}^2 = 1$ (because it generates
${\mathbb Z}_2$).
This new operator plays a role in the decomposition of the entire theory.

In terms of symmetries, 
\begin{itemize}
    \item The worldsheet has a global $B {\mathbb Z}_2$ (one-form) symmetry, from the fact that we are gauging a trivially-acting ${\mathbb Z}_2$ zero-form symmetry.
    \item In the spacetime we have a gauge(d) $B {\mathbb Z}_2$ (one-form) symmetry.
\end{itemize}
Again, this naturally matches the usual correspondence between global symmetries on the worldsheet and gauge symmetries in spacetime.

Since we have gauged a trivially-acting ${\mathbb Z}_2$ in the worldsheet
bulk (by virtue of the fact that the target is the gerbe $X_z$ \cite{Pantev:2005rh,Pantev:2005wj,Pantev:2005zs}), and since we have also gauged a trivial ${\mathbb Z}_2$ action on the
boundary, there is a decomposition \cite{Hellerman:2006zs} into two universes,
indexed by characters of ${\mathbb Z}_2$.  This is reflected by, for example, the bulk twist field for the trivially-acting ${\mathbb Z}_2$ generator, as well as the extra operators in the boundary theory.  
(This should be contrasted with the open strings described in e.g.~\cite{Hellerman:2006zs}, where we gauge a trivially-acting symmetry in bulk but not on the boundary.  There, the closed string theory decomposes, but the open string theory sees (for a fixed irreducible representation) only one of those universes, determined by the action on the boundary (the irreducible representation), so that a single open string theory, with a fixed action on the boundary defined by an irreducible representation,
does not itself decompose.)

More generally, if the boundary degrees of freedom have a global $H$ zero-form
symmetry, and we gauge a trivially-acting finite abelian group $K$ on bulk and
boundary, then we have a decomposition into $|K|$ universes, describing on the target space a decomposition into $|K|$ copies of Chern-Simons theory for $H$, indexed by characters of $K$.

\section{Proposal for general case}
\label{sect:genl}

In the previous two sections, we reviewed two special cases of worldsheet theories for gauging one-form symmetries in Chern-Simons theories:
\begin{itemize}
    \item We described
how the worldsheet for
the $G=H/Z$ ($Z$ is the center of $H$) Chern-Simons theory is a sigma model with target a $Z$ gerbe, described by a gauged $Z$ zero-form symmetry. Chan-Paton factors supported by the worldsheet boundary describing an $H$ bundle over that gerbe, which is also manipulated under the $Z$-gauging.
\item We
also described how the worldsheet realization of gauging a trivially-acting $BK$ in the Chern-Simons theory involves gauging a trivially-acting $K$ in the worldsheet bulk and boundary. This leads to a target-space theory given by a disjoint union of $|K|$ copies of Chern-Simons theory (possibly with different global structures for the $B$-field) for a given Lie group $H$.
\end{itemize}
In this section we combine these ingredients and generalize to the worldsheet described of target-space Chern-Simons for gauge group $H$, with $BA$ gauged, for finite abelian group $A$, as in 
\cite{Pantev:2022pbf}, for more general $A$ which may be neither completely in the center of $H$ nor purely trivially acting.

Let $Z$ be the image of $d: A \rightarrow H$, so that we have a short exact sequence
\begin{equation}  \label{eq:ses:kaz}
    1 \: \longrightarrow \: K \: \longrightarrow \: A \: \longrightarrow \: Z \: \longrightarrow \: 1
\end{equation}
of abelian groups.
Briefly,
we propose that the worldsheet realization of the gauged Chern-Simons theory has worldsheet given by a topologically twisted sigma model with target an $A$ gerbe (specified momentarily), in which we also gauge the action of $A$ on the boundary.  Locally, in other words, we gauge the action of $A$ on both bulk and boundary.
As a consequence, since $K \subset A$ acts trivially on both bulk and boundary, the worldsheet theory
has a global $BK$ (one-form) symmetry, and so decomposes into copies of the
$A/K = Z$ orbifold, indexed by irreducible representations of $K$
(as the trivially-acting subgroup is central) \cite{Hellerman:2006zs}.

Now, we shall be more precise.
Fix a $G=H/Z$ bundle $V$ on the target Lagrangian submanifold $M$, and let $w_G \in H^2(M,Z)$ be the obstruction to lifting the $G$ bundle to an $H$ bundle.  Let $a \in H^2(M,A)$ be a lift of $w_G$.  (We shall see momentarily that this lift only exists under special circumstances, which duplicate a target-space constraint discussed in \cite{Pantev:2022pbf}.)  Let $M_w$ be a $Z$-gerbe on $M$ of characteristic class $w_G$, and $M_a$ an
$A$-gerbe on $M$ of characteristic class $a$.  Let $X_a$ be the pullback of the $A$-gerbe $M_a \rightarrow M$ to $T^* M$, meaning that $X_a$ is an $A$-gerbe on $X = T^* M$.
Let $V_w$ be an $H$ bundle on $M_w$ (equivalently, a $Z$-twisted $H$ bundle on $M$) that projects to $V$, and let $V_a$ be its pullback to $M_a$.

We claim that the worldsheet realization of 
\begin{equation}
    \left[ {\rm Chern-Simons}(H) / BA \right]
\end{equation}
is a topologically twisted sigma model with target the $A$-gerbe $X_a$, describing a D-brane on $M_a \subset X_a$, where the boundary Chan-Paton factors couple to $V_a$, before gauging.
The action of $A$ on the boundary is then gauged.
(The trivial action of $A$ on the bulk is also gauged, implicit in the definition of a sigma model with target an $A$-gerbe \cite{Pantev:2005rh,Pantev:2005wj,Pantev:2005zs}.)
This clearly reduces to the two special cases described in the previous sections, and so generalizes both.

Now, there is a catch: not every $Z$ gerbe 
lifts to an $A$ gerbe.
The obstructions
lie in $H^3(M,K)$, and are the image of $w_G$ under the Bockstein homomorphism $\beta$ in
the long exact sequence associated
to~(\ref{eq:ses:kaz}):
\begin{equation}
H^2(M,A) \: \longrightarrow \: H^2(M,Z) \: \stackrel{\beta}{\longrightarrow} \:
H^3(M,K).
\end{equation}
As a result, the only $G$ bundles that can be described by this proposal
all have $\beta(w_G) = 0$, which is the worldsheet realization of the target-space constraint
described in \cite{Pantev:2022pbf}.
The fact that our construction naturally duplicates
this target-space constraint, is an important self-consistency check.

The reader may also observe that the choice of $A$-gerbe lifting $M_w$ is not unique, and in fact
the set of $K$-gerbes on $M$ acts on such choices.  This is a worldsheet representation of the target-space groupoid structure described in \cite{Pantev:2022pbf}.

Let us double-check that along the boundary, gauging $A$ has the effect of projecting onto $Z$-invariant states (and not, say, some projectivization thereof).
To see this, we recall (see appendix~\ref{app:gauge-qm}) that although there are no
twisted sectors in the boundary states, the effect of gauging $A$ is
to introduce a sum over boundary conditions along the timelike axis.
That sum over $A$ boundary conditions inserts a projector onto $A$-invariant
states.  More formally, if we write a projector\footnote{
It is easy to check that $P_A^2 = P_A$.
}
\begin{equation}  \label{eq:first-proj}
P_A \: = \: \frac{1}{|A|} \sum_{a \in A} \tau_a,
\end{equation}
where $\tau_a$ is an operator implementing the action of $A$,
then gauging $A$ on the boundary has the effect, via the sum over
boundary conditions, of inserting $P_A$ into the Wilson line on the
boundary.  Furthermore, since $K \subset A$ acts trivially on states,
\begin{eqnarray}
P_A | \psi \rangle & = &
\frac{1}{|A|} \sum_{a \in A} \tau_a | \psi \rangle,
\\
& = &
\frac{1}{|A|} \sum_{a \in A} \tau_{\pi(a)} | \psi \rangle,
\\
& = &
\frac{|K|}{|A|} \sum_{K-{\rm orbits}} \tau_{\pi(a)} | \psi \rangle,
\\
& = &
\frac{1}{|Z|} \sum_{x \in Z} \tau_x | \psi \rangle,
\\
& = & P_Z | \psi \rangle,
\end{eqnarray}
where $P_Z$ is the analogous projector onto $Z$-invariant states.
So, summing over $A$ boundary conditions
along the boundary Wilson line
is equivalent to inserting a projector onto $Z$-invariant states.

Finally, let us turn to the decomposition.  So far we have argued for a worldsheet realization of
the target-space one-form symmetry gauging
\begin{equation}
    \left[ {\rm Chern-Simons}(H) / BA \right].
\end{equation}
We can also see the claimed decomposition of the target-space theory.
Briefly, this follows from a worldsheet decomposition.
The finite abelian group $A$ is gauged in both bulk and boundary, but the 
normal subgroup $K \subset A$ acts trivially on both bulk and boundary.
As a result, since the two-dimensional theory has a gauged trivially-acting central symmetry\footnote{The central symmetry here denotes $K$ center of the gauged $A$ group, instead of the center $Z$ of the gauge group $H$.} ,
the worldsheet theory decomposes into universes indexed by characters of $K$.

Now, just as in \cite{Pantev:2022pbf}, in that decomposition,
the target-space restriction to $G$ bundles of
$\beta(w_G) = 0$ is implemented by summing over contributions from Chern-Simons theories with different discrete theta angles, each coupling to $\beta(w_G)$ but with a different character of $K$.
Gauging the trivial action of $K$ in the bulk results in a decomposition into universes indexed by characters of $K$, so we see that to be consistent, each of the Chern-Simons theories in that decomposition must also be weighted by a discrete theta angle, precisely in accord with
\cite{Pantev:2022pbf}.  

This analysis is much akin to the analysis of the example of the
$[X/D_4]$ orbifold theory in two dimensions with trivially-acting central ${\mathbb Z}_2$, discussed in
\cite[section 5.2]{Hellerman:2006zs}.  There, the partition function of the $D_4$ orbifold was shown to look like that of a ${\mathbb Z}_2 \times {\mathbb Z}_2$ orbifold but with missing sectors
(and an extra overall factor of 2), which could be understood as a sum of partition functions of 
orbifolds with and without discrete torsion.  The choice of discrete torsion there is a precise analogue of the target-space discrete theta angle here, arising when gauging the one-form $K$ center symmetry of the $H$ gauge theory.

In principle, it would be nice to also give a worldsheet description of the individual discrete theta angles appearing in target-space Chern-Simons theories, coupling to $\beta(w_G)$.
For example, it is natural to speculate that they can be understood in the same vein
as \cite{Freed:1999vc}, perhaps combined with the Green-Schwarz mechanism.
Later in section~\ref{sect:undoing} we shall see evidence that they should be realized by  $B$ fields.
It is also natural to speculate that if they can be understood in terms of $B$ fields with curvature,
whether the constraint $\beta(w_G) = 0$ is a discrete analogue of the old statement that in
a sigma model with nonzero $H$, the possible
D-branes are constrained (rank $r$ twisted bundles only exist if\footnote{
This can be seen explicitly as follows.
Consider bundles twisted by an ordinary nontrivial $B$ field in more detail.
The transition functions $g_{ab}$ of such a twisted bundle, of say rank $r$, obey
\begin{equation}
g_{ab} g_{bc} g_{ca} \: = \: \alpha_{abc},
\end{equation}
on triple overlaps,
where $[\alpha] \in H^2(X, C^{\infty}(U(1)))$.  Let
$\overline{g}_{ab}$ denote the image of the transition functions in
$PGL(r)$, then
on triple overlaps
\begin{equation}
\overline{g}_{ab} \overline{g}_{bc} \overline{g}_{ca} \: = \: 1,
\end{equation}
and $[ \overline{g} ] \in H^1(X, C^{\infty}( PGL(r) ) )$.  From the exact
sequence of sheaves
\begin{equation}
1 \: \longrightarrow \: {\mathbb Z}_r \: \longrightarrow \:
C^{\infty}( GL(r) ) \: \longrightarrow \:
C^{\infty}( PGL(r) ) \: \longrightarrow \: 1,
\end{equation}
there is the associated Bockstein homomorphism
\begin{equation}
\beta': \: H^1(X, C^{\infty}( PGL(r) ) ) \: \longrightarrow \: H^2(X,
{\mathbb Z}_r ),
\end{equation}
and it is straightforward to show that $\beta'( [\overline{g}] )$ coincides
with the image of $[\alpha]$ under an embedding of ${\mathbb Z}_r 
\rightarrow U(1)$.  Roughly, we could write $\beta'([\overline{g}]) = [\alpha]$,
and so we see mathematically
a rank $r$ twisted bundle only exists if $[\alpha]$ is $r$-torsion.
}
the curvature is $r$-torsion, see \cite[section 2]{Caldararu:2003kt} and
references therein, so if $H$ is nonzero and nontorsion, D-branes are
restricted to lie on loci where the restriction of $H$ vanishes,
as in e.g.~WZW models \cite{Klimcik:1996hp,Alekseev:1998mc}).
In any event, we leave the precise worldsheet description of discrete theta angles on different factors for future work.

\section{Example:  $SU(2)/{\mathbb Z}_4$ with trivially-acting kernel ${\mathbb Z}_2$}
\label{sect:su2-z4}

As a simple example, consider the special case that $H=SU(2)$, $A = {\mathbb Z}_4$, and
$K = Z = {\mathbb Z}_2$.  This case was discussed in the previous paper
\cite{Pantev:2022pbf}, where it was argued that
\begin{equation}
    \left[ {\rm Chern-Simons}(SU(2)) / B {\mathbb Z}_4 \right] \: = \:
    \coprod_{ \theta \in \hat{\mathbb Z}_2} {\rm Chern-Simons}(SO(3))_{\theta},
\end{equation}
where the subscript denotes discrete theta angles.
In this particular example, it was also argued in \cite{Pantev:2022pbf} that, at least for oriented $M$, the discrete theta angles all vanish\footnote{From a $3d$ field-theory perspective, this can be understood as follows. $SO(3)$ theories with different discrete theta angles can be derived from gauging one-form $\mathbb{Z}_2$ symmetry of $SU(2)$ theory, with and without discrete torsion. Schematically,
\begin{equation}
\begin{split}
	Z_{SO(3)_+}[B_1]&=\sum_{b_2}Z_{SU(2)}[b_2]e^{\pi i\int_{M} b_2\cup B_1 },\\
	Z_{SO(3)_-}[B_1]&=\sum_{b_2}Z_{SU(2)}[b_2]e^{\int_{M} \beta(b_2)}e^{\pi i\int_{M} b_2\cup B_1 }.
\end{split}
\end{equation}
For an oriented manifold $M$, $\beta(b_2)=0$, i.e., discrete torsion cannot be effectively turned on. This thus trivializes the discrete theta angle choices after gauging.}.

This is realized on the worldsheet as follows.
Let $V$ be a $SO(3)$ bundle on $M$.  Let $w_2$ be its second Stiefel-Whitney class (the obstruction to lifting to an $SU(2)$ bundle), and let $a \in H^2(M,{\mathbb Z}_4)$ be a lift of $w_2$.
(In general, this lift only exists if $\beta(w_2) = 0$; here, however, so long as $M$ is oriented,
$\beta(w_2) = 0$ always, so a lift $a$ always exists.)  Let $M_a$ be a ${\mathbb Z}_4$ gerbe on
$M$ of characteristic class $a$, $X_a$ a ${\mathbb Z}_4$ gerbe on $X = T^* M$ given as the
pullback of $M_a \rightarrow M$.  Let $V_w$ be a $SU(2)$ bundle on a ${\mathbb Z}_2$ gerbe on $M$ of characteristic class $w_2$ lifting $V$ (equivalently, a ${\mathbb Z}_2$-twisted $SU(2)$ bundle
on $M$ that projects to $V$), and let $V_a$ be its pullback to $M_a$, pulled back along the
projection $M_a \rightarrow M$.

Then, the proposed worldsheet description is that the bulk is a topologically-twisted sigma model with target the $A$-gerbe $X_a$,
with boundary along $M_a \subset X_a$ and Chan-Paton factors given by $V_a$.  This means that there is a (locally)
gauged ${\mathbb Z}_4$, all of which acts trivially on the bulk degrees of freedom, and 
a ${\mathbb Z}_2 \subset {\mathbb Z}_4$ acts trivially on boundary degrees of freedom.
Because there is a gauged central ${\mathbb Z}_2$ that acts trivially on everything, the worldsheet theory decomposes into two universes, each of which is separately a worldsheet theory for
$SU(2)/{\mathbb Z}_2 = SO(3)$ Chern-Simons as a string field theory.

\section{Gauging the target-space 2-form symmetry} 
\label{sect:undoing}

Having discussed how the decomposition of three-dimensional Chern-Simons theories can be realized via open string worldsheet theory, let us now discuss how the decomposition can be ``undone." Recall that decomposition in $d$-dimensional theories can be regarded as the existence of a $(d-1)$-form symmetry, thus undoing the decomposition, i.e., selecting one of the universes can be realized as gauging the $(d-1)$-form symmetry \cite{Sharpe:2019ddn}. In the case of $3d$ decomposed Chern-Simons theories,
\begin{equation}
    [{\rm Chern-Simons}(H) / BA] \: = \: \coprod_{\theta \in \hat{K}} {\rm Chern-Simons}(G)_{\theta},
\end{equation}
we should gauge the global two-form symmetry in the target-space theory responsible for the decomposition.

There is a very similar story for the worldsheet theory.  The different universes of the target-space theory also arise as different universes in the worldsheet theory, which has a global
$B K$ (one-form) symmetry.  Gauging that $BK$ on the worldsheet (as in \cite{Sharpe:2019ddn}) will select out one of the universes, and so return the worldsheet theory for a single Chern-Simons theory.

The universe selected when gauging the $(d-1)$-form symmetry is determined by a 
discrete theta angle for that symmetry.
Recall that when gauging a $p$-form finite symmetry, one needs to pick a background field profile for the quantum $(d-p-1)$-form symmetry. In the case of undoing the decomposition, this corresponds to choosing a parameter for the quantum $(-1)$-form symmetry, thus selecting a universe. This analysis suggests that the worldsheet realization of the target-space discrete theta angles
should be in terms of $B$ fields, as they index universes in related contexts in
\cite{Sharpe:2019ddn}.  We leave that question for future work.

\section{Analogues in the B model}   \label{sect:bmodel}

So far we have only discussed topologically A-twisted worldsheets.
However, as noted in \cite{Witten:1992fb}, there is an analogue for the B model.
Specifically, the classical open string field theory of the B model is holomorphic
Chern-Simons theory on the ambient Calabi-Yau X, rather than ordinary
Chern-Simons on a special Lagrangian submanifold $M \subset X$.

Now, the same worldsheet construction we have described can also be defined in the topological B model, and for the same reasons we have discussed, its string field theory will decompose.  However, since holomorphic Chern-Simons theory is a six-dimensional theory, not a three-dimensional theory, some of the target-space interpretations necessarily change.

We do not claim to have carefully checked holomorphic Chern-Simons, but the worldsheet physics suggests the following target-space picture.
In general terms,
the target-space interpretation of the B twist should be gauging a four-form symmetry in holomorphic Chern-Simons theory. The gauged four-form symmetry is then associated with the worldsheet global one-form symmetry, and in the case $K \neq 0$, since the worldsheet theory decomposes, we expect that the target-space theory decomposes, just as for the topological A twist and ordinary Chern-Simons.  
As holomorphic Chern-Simons is a six-dimensional theory, the existence of decomposition ordinarily requires that the target-space theory must admit a global five-form symmetry.  This is consistent if the target-space theory is
    \begin{equation}
       \left[ \mbox{Holomorphic Chern-Simons}(H) / B^4 A \right],
    \end{equation}
where $B^4 A$ acts analogously to the ordinary Chern-Simons case
(with $Z = {\rm im}\, d$ acting via the center of $H$, with trivially-acting
kernel $K$).  
In the case $K \neq 0$, as mentioned above, we expect a decomposition,
which suggests
\begin{equation}
    \left[ \mbox{Holomorphic Chern-Simons}(H) / B^4 A \right] \: = \: 
    \coprod_{\theta \in \hat{K}} \mbox{Holomorphic Chern-Simons}(G)_{\theta},
\end{equation}
for some choice of discrete theta angle,
the same pattern as for ordinary Chern-Simons described in \cite{Pantev:2022pbf}. 
On the left, since the gauged $B^4 K \subset B^4 A$ acts trivially,
the theory would have a global $B^5 K$ symmetry, consistent with the decomposition.
As before the worldsheet bulk theory has a global $BA$ symmetry,
which in this case would become a gauged $B^4A$ symmetry on the target space.

The description above seems most likely, but for completeness, we should mention that there might exist another possibility, in the form of a `transverse' decomposition along the lines of
\cite{Sharpe:2024ujm}, here in the holomorphic sector of the six-dimensional theory.

We leave a more detailed investigation of implications for holomorphic Chern-Simons to future work.

\section{Conclusions}

In this paper we have proposed a worldsheet description of (one-form symmetry) gauged Chern-Simons theories, a worldsheet
realization of the decomposition
of \cite{Pantev:2022pbf},
generalizing the picture of Chern-Simons as the open string field theory of the A model
\cite{Witten:1992fb}.

We speculated (in section~\ref{sect:bmodel}) about analogues 
in holomorphic Chern-Simons.  One may similarly speculate that the considerations here may have applications in topological strings, as in
\cite{Marino:2005sj}, which we leave for the future.

It would be interesting to understand applications via Gopakumar-Vafa's results  \cite{Gopakumar:1998ki}.  Their paper proposed a duality between large $N$ $SU(N)$ Chern-Simons on $S^3$ and topological closed string theory on a small resolution of the conifold.  Now,
the large $N$ limit of ${\mathbb Z}_N$ is $S^1$, so the
large $N$ limit of the center symmetry is, at least morally, $BU(1)$.
This suggests that gauging the center symmetry may be dual to a countably infinite collection of topological strings,
which could be identified with \cite{Cherman:2020cvw,Nguyen:2021yld,Nguyen:2021naa} a topological string theory trivially coupled to a free $U(1)$ gauge theory, so that the large $N$ limit of such a gauging is the original duality, times a free
$U(1)$ gauge theory.  We leave such developments for the future.

\section{Acknowledgements}

We would like to thank O.~Aharony, P.~Aspinwall, R.~Blumenhagen,
R.~Bryant, U.~Schreiber, S.~Sethi, Y.~Tachikawa, and E.~Witten
for useful conversations.  
T.P. was partially supported by NSF/BSF grant DMS-2200914, NSF FRG grant DMS 2244978, and by Simons HMS Collaboration grant 347070. 
E.S. was partially supported by NSF grant PHY-2310588. X.Y. was supported by NSF grant PHY-2014086.
X.Y. thanks the ICTP String-Math 2024, the Institute of Theoretical Physics (IFT), and the 2024 IHES Summer School - Symmetries and Anomalies: A modern take, for their hospitality during part of this work. 

\appendix

\section{Chern-Simons theory as A model string field theory}
\label{app:revcs}

In this section, we briefly review the realization of the Chern-Simons theory as an open string field theory of the topological A model to assist the reader.
We refer to \cite{Witten:1992fb} for details. 

Consider the topological A model, a sigma model of maps
\begin{equation}
	\Phi: \Sigma \rightarrow X
\end{equation}
from the worldsheet manifold $\Sigma$ to a target space geometry $X$. In this paper, we consider the special case where $X$ is a Calabi-Yau 3-fold, with Ricci-flat K\"{a}hler metric $g_{i\bar{j}}$. 

The bosonic fields give rise to local coordinates $\phi^I(x^\alpha)$ for the target space $X$, as functions of the worldsheet coordinate $x^\alpha, \alpha=1,2$. There are two classes of fermionic fields, denoted by $\chi^I$ and $\psi$. $\chi^I$ corresponds to sections of $\Phi^*(TX)$, while $\psi$ corresponds to a one-form valued in $\Phi^*(TX)$, where $TX$ is the tangent bundle of $X$. More precisely, the $\psi$ fields are associated to 
\begin{equation}
	\psi_z^{\bar{i}} \in \Phi^*(T^{0,1}X), ~\psi_{\bar{z}}^{i} \in \Phi^*(T^{1,0}X).
\end{equation}
The action,
\begin{equation}
	L=2t\int_{\Sigma}d^2z\left( \frac{1}{2}g_{IJ}\partial_z\phi^I\partial_{\bar{z}}\phi^J+i\psi_z^{\bar{i}}D_{\bar{z}}\chi^ig_{\bar{i}i}+ i\psi_{\bar{z}}^{i}D_{z}\chi^{\bar{i}}g_{\bar{i}i}-R_{i\bar{i}j\bar{j}}\psi^i_{\bar{z}}\psi^{\bar{i}}_z\chi^i\chi^{\bar{j}}\right),
\end{equation}
is BRST exact,
where the parameter $t$ plays a similar role as $1/\alpha^\prime$  in physical string theory.

We are interested in open string theory, i.e., worldsheets with boundary. Along any component of the boundary, we specify 
\begin{itemize}
    \item boundary conditions on bulk fields, defining a Lagrangian\footnote{One can also specify coisotropic subspaces, but for our purposes, we only consider Lagrangian cases.} submanifold $M\subset X$, 
    \item boundary degrees of freedom, defining a connection $A=A_Id\phi^I$ with structure group $G$ on a vector bundle $E$ over $M$. 
\end{itemize}

The quantum Hilbert space $\mathcal{H}$ (in the large $t$ limit) can be built from the canonical commutation relations $[\frac{d\phi^I}{d\tau}(\sigma), \phi^J(\sigma^\prime)]=-\frac{i}{t}g^{IJ}\delta(\sigma-\sigma^\prime)$ and $\{ \psi_\tau(\sigma), \chi(\sigma^\prime) \}=\frac{1}{t}\delta(\sigma-\sigma^\prime)$. 
Functionals $\mathcal{A}$ in $\mathcal{H}$ can be expanded by zero-modes\footnote{Zero-modes for $\psi$ fields do not appear because they can be expressed as $\psi_\tau^a\leftrightarrow \frac{\partial}{\partial \chi^a}$, due to the canonnical conjugation.}
\begin{equation}\label{eq: expansion of open string functional}
	\mathcal{A}(q^a,\chi^a)=c(q)+\chi^aA_a(q)+\chi^a\chi^bB_{ab}(q)+\cdots
\end{equation}
where $c, A_a, B_{ab}$ are $p$-forms on the Lagrangian submanifold $M$ with $p\in \{ 0,1,2,3 \}$. Including the information on the boundary connections with structure group $G$, i.e., information of Chan-Paton factors, these target space $p$-forms are valued in the endomorphisms End$(E)$ of the vector bundle $E$ over $M$. The nilpotent BRST operator $Q$ with (anti-)commutators $[Q, \phi^I]=-\chi^I$ and $\{ Q, \chi^J \}=0$ has a target space interpretation as the exterior derivative $Q\leftrightarrow d$ on $M$, constructing the cohomology $H^*(M, \text{End}(E))$.

Let us now move to the string field theory. The necessary information we need is a $\mathbb{Z}$-graded associated algebra $\mathcal{B}$, with multiplication $\star$, a nilpotent derivation operator $Q$ with ghost number 1 (or equivalently, degree 1), and a linear functional $\int: \mathcal{B}\rightarrow \mathbb{C}$ of ghost number $-3$. For a string field $\mathcal{A}\in \mathcal{B}$, consider the action 
\begin{equation}\label{eq: SFT action}
	L=\frac{1}{2}\int\left( \mathcal{A}\star Q\mathcal{A}+\frac{2}{3}\mathcal{A}\star\mathcal{A}\star \mathcal{A} \right).
\end{equation} 
For open strings, we include Chan-Paton factors labeled by rank-$N$ matrices $M_N$, and replace 
\begin{equation}
	\mathcal{B}\rightarrow \mathcal{B}\otimes M_N, ~\int \rightarrow \int \otimes \text{Tr}
\end{equation}

Since the ghost number of the linear action $\int$ is $-3$, the string field theory above needs the field $\mathcal{A}$ to have ghost number $1$. Therefore, among the states in open string Hilbert space $\mathcal{H}$, we require the expansion in (\ref{eq: expansion of open string functional}) to reduce to
\begin{equation}
	\mathcal{A}=\chi^aA_a(q),
\end{equation}
which is a one-form on $M$ valued in End$(E)$. In the large $t$ limit (where instanton corrections are suppressed), we have $Q\rightarrow d$, so the first term in (\ref{eq: SFT action}) reduces to 
\begin{equation}
	\frac{1}{2}\int_{M}\mathcal{A}\star Q\mathcal{A}\longrightarrow \frac{1}{2}\int_M\text{Tr}A\wedge dA,
\end{equation}
which is the single derivative term of the Chern-Simons theory. 
The cubic term in (\ref{eq: SFT action}) can be viewed as a three-point function for vertex operators $V^{(i)}=\chi^aA_a^{(i)}(q)$, where the $A^{(i)}$ are modes of $A$,
\begin{equation}
	\langle V^{(1)}(0)V^{(2)}(1)V^{(3)}(\infty) \rangle.
\end{equation}
In the large $t$ limit, this correlation function is just the integral over zero-modes
\begin{equation}
	\int dq^1\cdots dq^3d\chi^1\cdots d\chi^3\text{Tr}\chi^aA_a^{(1)}(q)\chi^bA_b^{(2)}(q)\chi^cA_c^{(3)}(q)=\int_M\text{Tr}A^{(1)}\wedge A^{(2)} \wedge A^{(3)},
\end{equation}
which exactly matches the cubic term for the conventional Chern-Simons theory.

\section{Open string gauge groups}
\label{app:global}

In this appendix, we briefly review standard results on gauge groups appearing in open strings, as their structure
is important for this paper.

At a perturbative level, the possible open string gauge algebras
are $\mathfrak{u}(n)$, $\mathfrak{so}(n)$, and $\mathfrak{usp}(2n)$ 
\cite{Paton:1969je,Schwarz:1982md,Marcus:1982fr}, realized by quantum mechanical systems on worldsheet boundaries with various parity operations imposed.  
The construction of the perturbative open
string states
at all mass levels was discussed in 
\cite{Paton:1969je,Schwarz:1982md,Marcus:1982fr} (see 
e.g.~\cite[section 1.3]{Schwarz:1982jn}, 
\cite{Marcus:1986cm,Weinberg:1987ie,Bianchi:1990yu,Angelantonj:2002ct,Polchinski:1996fm}.
Briefly, the perturbative open string states  are all
in the singlet, fundamental symmetric 2-tensor, and antisymmetric 2-tensor
representations (and to get a fundamental, one needs an open string ending
on a different D-brane). 
For example, from  \cite[table 1]{Marcus:1982fr}, \cite{Marcus:1986cm},
\begin{itemize}
\item For $\mathfrak{so}(n)$, the even-mass-level states are in the
adjoint (antisymmetric 2-tensor) representation, and the odd-mass-level
states are in the traceless symmetric 2-tensor and singlet representations.
\item For $\mathfrak{usp}(2n)$, the even-mass-level states are in the adjoint (symmetric 2-tensor)
and the odd-mass-level states are in the
traceless antisymmetric and singlet representations.
\end{itemize}

Now, nonperturbatively, there can be additional states.
For example, perturbative heterotic $Spin(32)/{\mathbb Z}_2$
strings have massive states in
additional representations, so duality with the type I string
implies those states must arise as some sort of solitonic states.
In particular, \cite[section 5]{Witten:1995ex},
\cite[section 2]{Witten:1998cd}  observes that the heterotic 
Spin$(32)/{\mathbb Z}_2$
theory has particles that transform as spinors, which are absent in the
perturbative type I spectrum, and so must arise as solitons.
(See also \cite{Polchinski:1995df} for related observations regarding
the relation between heterotic and type I string states.)

In particular, the perturbative mass spectrum constrains possible global
forms of the gauge group, but does not uniquely determine the global
form of the gauge group, which must be determined independently via nonperturbative effects.

For our purposes in this paper, we work abstractly with boundary quantum mechanical systems with some global symmetry, which determines the Chern-Simons gauge symmetry on the target space.

\section{Gauging in quantum mechanics}  \label{app:gauge-qm}

In this appendix, we briefly discuss gauging a finite group $G$ in quantum mechanical ($0+1$ dimensional) systems, and compare and contrast with two-dimensional orbifolds.

In two dimensions, because the spatial cross-section is a circle, there exist twisted sectors, extra contributions to the Hilbert space\footnote{This is sometimes referred to as defect Hilbert space in more recent literature.}.  By contrast, in quantum mechanics, since a spatial cross section is just a point, there are no twisted sectors, no analogous additional contributions to the Hilbert space.

That said, the quantum-mechanical gauge theory will contain extra operators, although it will not have extra states.  This is because the $(0+1)$-dimensional path integral sums over $G$ bundles, just as happens in any other dimension when gauging a finite group, and those bundles can be nontrivial.  If for example $G$ is abelian, then equivalence classes of $G$ bundles on $S^1$ are in one-to-one correspondence with elements of $G$.  Those different bundles can be equivalently represented by adding new operators, analogous to twist fields (but only creating boundary conditions, not generating twisted sectors), and because the path integral sums over bundles, any partition function will sum over those operator insertions, which has the effect of inserting a projector -- so only $G$-invariant states propagate in the quantum-mechanical theory.  If $G$ acts trivially, one still gets new operators (though not new states).

From a generalized symmetry perspective, an ordinary $G$ symmetry is a $(d-1)$-form symmetry for a quantum mechanical system, thus any quantum mechanical system with a global (zero-form) symmetry decomposes. This aligns with the fact that if a quantum mechanical system is $G$-symmetric, the theory has superselection sectors (now understandable as universes). The projectors discussed above are built from the topological local operators generating the $G$ symmetry for the quantum mechanics. Gauging the $G$ symmetry is then translated into summing over topological local operators and picks one of the universes (superselection sectors) of the quantum mechanical system. 

The above statement can be generalized easily. If a $d$-dimensional theory
has a global $(d-1)$-form symmetry, it decomposes.
Gauging the global $(d-1)$-form symmetry selects out one universe in the decomposition, and yields a theory with a quantum $(-1)$-form symmetry. The background field for the $(-1)$-form symmetry is just the parameter labeling the universe in the decomposition.  Gauging the quantum $(-1)$-form symmetry then sums over universes and returns the original theory with the global $(d-1)$-form symmetry (now a quantum symmetry of the $(-1)$-form symmetry gauging).

\end{document}